\documentclass[aps,pra,superscriptaddress,floatfix,showpacs,10pt, twocolumn]{revtex4}
\usepackage{bbm}
\usepackage{amsfonts}
\usepackage[dvips]{graphicx}
\usepackage{amsmath,amsfonts,amssymb,graphics,graphicx,epsfig,color,times,bbm}
\usepackage{longtable,subfigure}

\begin{document}

\title{Optimal reconstruction of the states in qutrits system}
\author{Fei Yan}
\affiliation{Key Laboratory of Opto-electronic Information
Acquisition and Manipulation, Ministry of Education, School of
Physics and Material Science, Anhui University, Hefei 230039,
People's Republic of China}
\author{Ming Yang\footnote{mingyang@ahu.edu.cn}}
\affiliation{Key Laboratory of Opto-electronic Information
Acquisition and Manipulation, Ministry of Education, School of
Physics and Material Science, Anhui University, Hefei 230039,
People's Republic of China}
\author{Zhuo-Liang Cao}
\affiliation{Department of Physics {\&} Electronic Engineering,
Hefei Normal University, Hefei 230061, People's Republic of China}

\begin{abstract}
Based on mutually unbiased measurements, an optimal tomographic
scheme for the multiqutrit states is presented explicitly. Because
the reconstruction process of states based on mutually unbiased
states is free of information waste, we refer to our scheme as the
optimal scheme. By optimal we mean that the number of the required
conditional operations reaches the minimum in this tomographic
scheme for the states of qutrit systems. Special attention will be
paid to how those different mutually unbiased measurements are
realized; that is, how to decompose each transformation that
connects each mutually unbiased basis with the standard
computational basis. It is found that all those transformations can
be decomposed into several basic implementable single- and
two-qutrit unitary operations. For the three-qutrit system, there
exist five different mutually unbiased-bases structures with
different entanglement properties, so we introduce the concept of
physical complexity to minimize the number of nonlocal operations
needed over the five different structures. This scheme is helpful
for experimental scientists to realize the most economical
reconstruction of quantum states in qutrit systems.
\end{abstract}

\pacs{03.65.Wj, 03.65.Ta, 03.65.Ud, 03.67.Mn}
\maketitle

\section{Introduction}
The quantum state of a system is a fundamental concept in quantum
mechanics, and a quantum state can be described by a density matrix,
which contains all the information one can obtain about that system.
A main task for implementing quantum computation is to reconstruct
the density matrix of an unknown state, which is called quantum
state reconstruction or quantum state
tomography\cite{Dariano03,Dariano04}. The technique was first
developed by Stokes to determine the polarization state of a light
beam \cite{GCStokes}. Recently,
 Minimal qubit tomography process has been proposed by \v{R}eh\'{a}\v{c}ek \emph{et al}, where only
four measurement probabilities are needed for fully determining a
single qubit state, rather than the six probabilities in the
standard procedure \cite{JR}. But the implementation of this
tomography process requires measurements of N-particle
correlations\cite{Ling}.The statistical reconstruction of biphotons
states based on mutually complementary measurements has been
proposed by Bogdanov \emph{et al}\cite{YIB1,YIB}. Ivanov \emph{et
al} proposed a method to determine an unknown mixed qutrit state
from nine independent fluorescence signals\cite{PAI}. Moreva
\emph{et al} paid attention to experimental problem of the
realization of the optimal protocol for polarization ququarts state
tomography \cite{EVM}. In 2009, Taguchi \emph{et al} developed the
single scan tomography of spatial three-dimensional (qutrits) state
based on the effect of realistic measurement operators\cite{GT}.
Allevi \emph{et al} studied the implementation of the reconstruction
of the Wigner function and the density matrix for coherent and
thermal states by by switching on/off single photon avalanche
photodetectors\cite{AAA}.

In order to obtain the full information about the system we need to
perform a series of measurements on a large number of identically
prepared copies of the system. These measurement results are not
independent of each other, so there is redundancy in these results
in the previously used quantum tomography processes \cite{RTT},
which causes a resources waste. If we remove this redundancy
completely, the reconstruction process will become an optimal one.
So, to design an optimal set of measurements for removing the
redundancy is of fundamental significance  in quantum information
processing.

Mutually unbiased bases (MUBs) have been used in a variety of topics
in quantum mechanics
\cite{IDI,WKW,SB,Revzen,JL,SC,AK,ABK,JLR,MP,TD,AOP,ABC,
WW,AJS,ABJ,IB,MAJ,SBS,MRK,PJ,SBr,AA,DMA}. MUBs are defined by the
property that the squared overlap between a vector in one basis and
all basis vectors in the other bases are equal. That is to say the
detection over a particular basis state does not give any
information about the state if it is measured in another basis.
Ivanovi\'{c} first introduced the concept of MUBs to the problem of
quantum state determination \cite{IDI}, and proved the existence of
such bases in the prime-dimension system by an explicit
construction. Then it has been shown by Wootters and Fields that
measurements in this special class of bases, i.e. mutually unbiased
measurements (MUMs) provide a minimal as well as optimal way of
complete specification of an unknown density matrix \cite{WKW}. They
proved that the maximal numbers of MUBs is $d+1$ in prime-dimension
system. This result also applies to the prime-power-dimension
system.

MUBs play a special role in determining quantum states, such as it
forms a minimal set of measurement bases and provides an optimal way
for determining a quantum state \cite{WKW, IDI, SB, Revzen} etc.
Recently an optimal tomographic reconstruction scheme was proposed
by Klimov \emph{et al} for the case of determining a state of
multiqubit quantum system based on MUMs in trapped ions system
\cite{PRA 77}. However, the use of three-level systems instead of
two-level systems has been proven to be securer against a symmetric
attack on a quantum key distribution protocol with MUMs than the
currently existing measurement protocol \cite{Bruss,Cerf}. Quantum
tomography in high dimensional (qudit) systems has been proposed and
the number of required measurements is $d^{2n}-1$ with $d$ being the
dimension of the qudit system and $n$ being the number of the qudits
\cite{RTT}. This tomography process is not an optimal one, and there
is a big redundancy among the measurement results there. To remove
this redundancy, we will propose an optimal tomography process for
qutrits states. This optimal quantum tomography process is the
MUBs-based qutrit states tomography, and the number of required
measurements is greatly reduced. A $d$-dimensional quantum system is
represented by a positive semidefinite Hermitian matrix $\rho$ with
unit trace in $d$-dimensional Hilbert space, which is specified by
$d^{2}-1$ real parameters. A nondegenerate measurement performed on
such a system provides $d-1$ independent probabilities. So, in
general, one requires at least $d + 1$ different orthogonal
measurements for fully determining an unknown $\rho$. For $n$-qutrit
tomography it only needs $3^{n}+1$ measurements. Through analysis,
one can find that the tomography process for qubit system can not be
generalized to qudit system case in a trivial way. The MUBs-based
tomography process for qubit system proposed by Klimov \emph{et al}
can not be directly applied to qutrit system \cite{PRA 77}. This is
because the entanglement feature of the MUBs of qubits system is
totally different from that of qutrits system. So, we will study the
physical implementation of an optimal tomographic scheme for the
case of determining the states of multi-qutrit system based on the
MUMs.

From the experimental point of view, the physical complexity is a
key point for the implementation of a scheme. Here, for multi-qutrit
quantum tomography, the physical complexity mainly comes from the
entanglement bases, i.e. the number of the two-qutrit conditional
operation needed for the decomposition of these entangled bases.  In
addition, there exists many different MUBs with different
entanglement properties in multi-qutrit system. So the physical
complexity of the quantum tomography process here depends on the
entanglement structure of the MUBs used, and it becomes very
important to optimize the quantum tomography process over all the
possible MUBs entanglement structures of the system.

This paper is arranged as follows. In the next section we introduce
the MUBs in a $d$-dimensional system ($d=p^{n}$ $(p\neq2)$) and show
how to reconstruct an unknown state by MUMs. Here $p$ is a prime.
Section \ref{single qutrit} briefly reviews the general method to
reconstruct a qutrit state, where the number of measurements is $8$.
However the MUBs-based qutrit tomography proposed here only needs
$4$ measurements, which means the number of measurements is reduced.
Here the measurement reduction for single qutrit case is not
obvious, so in section \ref{two-qutrit} we will extend the
one-qutrit case to two-qutrit system. For the two-qutrit system the
number of measurements is only $10$ for determining all the elements
of the density operator rather than $3^{4}-1=80$ measurements in the
scheme proposed in Ref. \cite{RTT}. It means a great reduction of
the experimental complexity. So we conclude that the optimal
measurements on the unknown qutrit states are the MUMs. In section
\ref{complexity} we discuss the physical complexity for implementing
the MUMs in three-qutrit system, and give the optimal MUB for the
qutrit system quantum tomography process with minimized physical
complexity. The last section is the conclusion.

\section{MUTUALLY UNBIASED BASES AND MUTUALLY UNBIASED MEASUREMENTS}\label{II}

As shown by Wootters, Fields \cite{WKW} and Klappenecker,
R\"{o}tteler \cite{AK}, in finite field language, the first MUB in a
$d=p^{n}$ $(p\neq2)$-dimensional quantum system is the standard
basis $B^{0}$ given by the vector
$(a_{k}^{(0)})_{l}=\delta_{kl},k,l\in\digamma_{p^{n}}$, where the
superscript denotes the basis, $k$ the vector in the basis, $l$ the
component and $\digamma_{p^{n}}$ is the field with $p^{n}$ elements.
The other $d$ MUBs are denoted by $B^{r}$ which consists of vectors
$(a_{k}^{(r)})_{l}$ defined by\cite{WKW}:
$(a_{k}^{(r)})_{l}=(1/\sqrt{d})\omega^{\mathcal {T}\emph{r}(r\cdot
l^{2}+k\cdot l)},r,k,l\in\digamma_{p^{n}},r\neq0.$
Here $\omega=exp(2\pi i/p)$ and
$\mathcal{T}\emph{r}\theta=\theta+\theta^{p}+\theta^{p^{2}}+\cdot\cdot\cdot+\theta^{p^{n-1}}.$
The set of mutually unbiased projectors can be given by
$P_{k}^{(r)}=|a_{k}^{(r)}\rangle\langle a_{k}^{(r)}|.$
It is worth noticing that $|a_{k}^{(r)}\rangle$ contains the
computational basis $B^{0}$. Here
$Tr(P_{j}^{(s)}P_{k}^{(r)})=(1/d)(1-\delta_{sr}+d\delta_{sr}\delta_{jk}).$
Then the measurement probabilities given by
$\omega_{k}^{(r)}=Tr(P_{k}^{(r)}\rho)$
completely determine the unknown density operator of a
$d$-dimensional system \cite{IDI}:
$\rho=\Sigma_{r=0}^{d}\Sigma_{k=0}^{d-1}\omega_{k}^{(r)}P_{k}^{(r)}-I.$

For instance, in a qutrit system, there are three MUBs beside the
computational basis $B^{0}=\{|0\rangle,|1\rangle,|2\rangle\}$, in
the following form with $\omega=exp(2\pi i/3)$:
\begin{subequations}
\begin{eqnarray}\label{a}
B^{1}&:&\{|a_{0}^{(1)}\rangle=(1/\sqrt{3})(|0\rangle+|1\rangle+|2\rangle),\nonumber\\&&
|a_{1}^{(1)}\rangle=(1/\sqrt{3})(|0\rangle+\omega|1\rangle+\omega^{*}|2\rangle),\nonumber\\&&
|a_{2}^{(1)}\rangle=(1/\sqrt{3})(|0\rangle+\omega^{*}|1\rangle+\omega|2\rangle)\};
\end{eqnarray}
\begin{eqnarray}\label{b}
B^{2}&:&\{|a_{0}^{(2)}\rangle=(1/\sqrt{3})(\omega|0\rangle+|1\rangle+|2\rangle),\nonumber\\&&
|a_{1}^{(2)}\rangle=(1/\sqrt{3})(|0\rangle+\omega|1\rangle+|2\rangle),\nonumber\\&&
|a_{2}^{(2)}\rangle=(1/\sqrt{3})(|0\rangle+|1\rangle+\omega|2\rangle)\};
\end{eqnarray}
\begin{eqnarray}\label{c}
B^{3}&:&\{|a_{0}^{(3)}\rangle=(1/\sqrt{3})(\omega^{*}|0\rangle+|1\rangle+|2\rangle),\nonumber\\&&
|a_{1}^{(3)}\rangle=(1/\sqrt{3})(|0\rangle+\omega^{*}|1\rangle+|2\rangle),\nonumber\\&&
|a_{2}^{(3)}\rangle=(1/\sqrt{3})(|0\rangle+|1\rangle+\omega^{*}|2\rangle)\}.
\end{eqnarray}
\end{subequations}

\section{RECONSTRUCTION PROCESS FOR AN ARBITRARY SINGLE QUTRIT
STATE}\label{single qutrit}
An unknown single qutrit state can be
expressed as \cite{RTT, James}:
$\rho=(1/3)\sum_{j=0}^{8}r_{j}\lambda_{_{j}},$
where $\lambda_{_{0}}$ is an identity operator and the other
$\lambda_{_{j}}$ are the SU(3) generators \cite{CM}. The general
method to reconstruct the qutrit state is to measure the expectation
values of the $\lambda$ operators \cite{RTT}, where
$r_{j}=\langle\lambda_{_{j}}\rangle=Tr[\rho\lambda_{_{j}}]$. Thus
one will find that the number of required measurements is $8$.
However, if we choose the MUMs to determine the qutrit state, the
number of needed MUMs is only $4$ rather than $8$ of Ref.
\cite{RTT}. The four optimal set of MUBs have been presented by
Eqs.(\ref{a},\ref{b},\ref{c}) plus the standard computational basis
in the preceding section. Each of the three MUBs in
Eqs.(\ref{a},\ref{b},\ref{c}) is related with the standard
computational basis by a unitary transformation. These
transformations have been listed in Table.\ref{Table.I.}. Here, $F$
denotes the Fourier transformation:
\begin{equation}
F|j\rangle=(1/\sqrt{3})\sum_{l=0}^{2}exp(2\pi
ilj/3)|l\rangle,j=0,1,2,
\end{equation}
$R$ denotes a phase operation:
\begin{equation}
R=|0\rangle\langle0|+\omega|1\rangle\langle1|+\omega|2\rangle\langle2|,
\end{equation}
and the Controlled gate is
\begin{equation}
X|i\rangle|j\rangle =|i\rangle |j\ominus i\rangle.
\end{equation}
Where $\ominus$ denotes the difference $j-i$ modulo 3.If there are
$n$ qutrits, the number of MUM is $3^{n}+1$, which is far less than
$3^{2n}-1$ in Ref. \cite{RTT}. That is to say the use of MUMs can
represent a considerable reduction in the operations and time
required for performing the full state determination \cite{PRA 77}.
\begin{table}[tbp]
\caption{The transformations connecting the MUBs with the standard
computational basis for a qutrit system based on Fourier transforms
and the phase operations.} \label{Table.I.}
\begin{ruledtabular}
\begin{tabular}{r@{}l}
\hline
Basis & Transformation\\
\hline
2 & $F^{-1}$ \\
3 &  $F^{-1}R$\\
4 &  $F^{-1}R^{-1}$\\
 \hline
\end{tabular}
\end{ruledtabular}
\end{table}

\section{RECONSTRUCTION PROCESS FOR AN ARBITRARY TWO-QUTRIT
STATE}\label{two-qutrit}
Now if we further extend one-qutrit case to
two-qutrit case, the density matrix can be expressed as:
$\rho_{12}=(1/9)\sum_{j,k=0}^{8}r_{jk}\lambda_{_{j}}\otimes\lambda_{_{k}},$
where $r_{jk}=\langle\lambda_{_{j}}\otimes\lambda_{_{k}}\rangle$. If
we use the general method in Ref. \cite{RTT} to fully determine the
state, $d^{2n}-1=3^{4}-1=80$ measurements will be needed. So much
measurements will inevitably introduce redundant information of the
state, which is obviously a resource waste. So here we will take
advantage of the MUMs to reconstruct the two-qutrit state. It is
easy to find that the nine elements of $\digamma_{9}$ (finite field)
are $\{0,\alpha,2\alpha,1,1+\alpha,1+2\alpha,2,2+\alpha,2+2\alpha\}$
by using the irreducible polynomials $\theta^{2}+\theta+2=0$
\cite{WKW}. Here we use the representation $\{|0\rangle,
|\alpha\rangle, |2\alpha\rangle \cdot\cdot\cdot |2+2\alpha\rangle\}$
as the standard basis.
\begin{table}[tbp]
\caption{The decompositions of MUBs for the two-qutrit system based
on Fourier transformations, phase operations and controlled-NOT
gates ($X_{12}$) \cite{ABR} with the first particle as source and
the second one as target. The subscript denotes the $i$th particle,
$i=1,2$.} \label{Table.II.}
\begin{ruledtabular}
\begin{tabular}{r@{}l}
\hline
Basis & Decompositions\\
\hline
2 & $F_{1}^{-1}F_{2}^{-1}$ \\
3 & $F_{1}^{-1}R_{1}F_{2}^{-1}R_{2}$ \\
4 &  $F_{1}^{-1}X_{12}F_{2}^{-1}R_{2}^{-1}$\\
5 &  $F_{1}^{-1}X^{-1}_{12}R_{1}F_{2}^{-1}R_{2}^{-1}$\\
6 &  $F_{1}^{-1}X^{-1}_{12}F_{2}^{-1}R_{1}^{-1}$\\
7 &  $F_{1}^{-1}R_{1}^{-1}F_{2}^{-1}R_{2}^{-1}$ \\
8 & $F_{1}^{-1}X^{-1}_{12}F_{2}^{-1}R_{2}$ \\
9 & $F_{1}^{-1}R_{1}^{-1}X_{12}F_{2}^{-1}R_{2}$ \\
10 & $F_{1}^{-1}R_{1}X_{12}F_{2}^{-1}$\\
 \hline
\end{tabular}
\end{ruledtabular}
\end{table}

One can find that there will be only $d^{2}+1=3^{2}+1=10$ MUMs to be
done, which is much less than $80$ of Ref. \cite{RTT}. It means that
the operations and time needed for the whole state determination is
great reduced. The decompositions for all the MUBs of the two-qutrit
system have been listed in Table. \ref{Table.II.}.

\section{THE PHYSICAL COMPLEXITY FOR IMPLEMENTING THE MUMS IN THE THREE-QUTRIT
SYSTEM}\label{complexity}

In general, the fidelity of single logic gates can be greater than
$99\%$, but nonlocal gates have a relatively lower fidelity. The
fidelity of a practical CNOT gate can reach a value up to 0.926 for
trapped ions system in Lab \cite{MR}. Klimov \emph{et al} have
introduced the concept of the physical complexity of each set of
MUBs as a function of the number of nonlocal gates needed for
implementing the MUMs \cite{PRA 77}. Here the fidelity value of the
CNOT gates for qubits systems also can be used to evaluate the
physical complexity of the MUBs of qutrits systems. Why we can say
so is because of the following point. Although the systems involved
here are three-state ones, all the operations used in our
reconstruction process can be decomposed into effective two-state
operations. So the complexity of the current tomography scheme is
proportional to the number of the nonlocal gates used($C\propto 6$)
for two qutrits system. As shown in Ref\cite{012302}, the only MUB
structure for a two-qutrit system is $(4,6)$, where $4$ is the
number of the separable bases and $6$ is the number of the bipartite
entangled bases.

However in three-qutrit case, there are five sets of MUBs with
different structures, namely $
\{(0,12,16)$,$(1,9,18)$,$(2,6,20)$,$(3,3,22)$,$(4,0,24)\}$
\cite{012302,JLR}. It is easy to see that the $(0,12,16)$ set of
MUBs has the minimum physical complexity. We say that the optimal
set of the MUBs is (0,12,16). The decompositions of the MUBs in the
three-qutrit case for (0,12,16) are listed in
Table.\ref{Table.III.}. So the set of MUBs (0,12,16) has a
complexity $C\propto 44$, which is a very important value in
experimental realization of it.

\begin{table*}[t]
\caption{The decompositions of MUBs (0,12,16) for three-qutrit
system} \label{Table.III.}
\centering\subtable{\begin{tabular}{lr} \hline
Basis & Decompositions\\
\hline
1 & $F_{2}^{-1}X_{23}F_{1}^{-1}F_{3}^{-1}$ \\
2 & $F_{1}^{-1}R_{1}^{-1}F_{2}^{-1}R_{2}^{-1}X_{23}F_{3}^{-1}R_{3}^{-1}$ \\
3 & $F_{1}^{-1}X_{13}R_{3}$ \\
4 & $F_{2}^{-1}R_{2}X_{23}^{-1}F_{1}^{-1}X_{12}R_{2}F_{3}^{-1}$\\
5 & $F_{1}^{-1}X_{12}^{-1}F_{3}^{-1}$\\
6 & $F_{2}^{-1}R_{2}X_{23}F_{1}^{-1}R_{1}X_{13}^{-1}R_{3}F_{3}^{-1}R_{3}$\\
7 & $F_{2}^{-1}X_{23}^{-1}R_{3}^{-1}F_{1}^{-1}R_{1}^{-1}X_{12}^{-1}R_{3}^{-1}$\\
8 & $F_{1}^{-1}R_{1}^{-1}X_{13}^{-1}F_{2}^{-1}R_{2}F_{3}^{-1}R_{3}^{-1}$\\
9 & $F_{1}^{-1}X_{13}^{-1}F_{2}^{-1}R_{2}X_{23}^{-1}F_{3}^{-1}R_{3}$\\
10 & $F_{1}^{-1}R_{2}X_{12}X_{13}^{-1}F_{2}^{-1}$\\
11 & $F_{1}^{-1}R_{1}X_{12}^{-1}F_{2}^{-1}R_{2}X_{23}^{-1}F_{3}^{-1}$\\
12 & $F_{1}^{-1}R_{1}^{-1}X_{13}X_{12}^{-1}F_{2}^{-1}F_{3}^{-1}R_{3}^{-1}$\\
13 & $F_{1}^{-1}X_{12}F_{2}^{-1}R_{2}^{-1}X_{23}^{-1}F_{3}^{-1}R_{3}^{-1}$\\
14 & $F_{1}^{-1}R_{1}X_{12}F_{2}^{-1}X_{23}F_{3}^{-1}R_{3}$\\
\hline
\end{tabular}}
\qquad \subtable{\begin{tabular}{lr}\hline
Basis & Decompositions\\
\hline
15 & $F_{2}^{-1}R_{2}X_{23}$\\
16 & $F_{1}^{-1}R_{1}X_{13}F_{2}^{-1}R_{2}X_{23}F_{3}^{-1}R_{3}^{-1}$\\
17 & $F_{1}^{-1}R_{1}^{-1}X_{13}F_{2}^{-1}F_{3}^{-1}$\\
18 & $F_{1}^{-1}R_{1}^{-1}X_{13}^{-1}F_{2}^{-1}R_{2}^{-1}X_{23}^{-1}R_{3}F_{3}^{-1}R_{3}^{-1}$\\
19 & $F_{1}^{-1}R_{1}X_{13}^{-1}F_{2}^{-1}R_{2}^{-1}F_{3}^{-1}R_{3}$\\
20 & $F_{1}^{-1}X_{12}F_{3}^{-1}R_{3}^{-1}$\\
21 & $F_{1}^{-1}R_{1}^{-1}X_{12}^{-1}F_{2}^{-1}R_{2}F_{3}^{-1}R_{3}$\\
22 & $F_{1}^{-1}X_{12}^{-1}F_{2}^{-1}X_{23}^{-1}F_{3}^{-1}R_{3}$\\
23 & $F_{1}^{-1}R_{1}^{-1}X_{13}F_{2}^{-1}R_{2}^{-1}X_{23}^{-1}$\\
24 & $F_{1}^{-1}R_{1}^{-1}X_{13}^{-1}F_{2}^{-1}X_{23}R_{3}F_{3}^{-1}$\\
25 & $F_{1}^{-1}R_{1}X_{12}^{-1}X_{23}^{-1}F_{2}^{-1}R_{2}^{-1}$\\
26 & $F_{1}^{-1}R_{1}X_{12}X_{23}^{-1}F_{2}^{-1}R_{2}^{-1}F_{3}^{-1}$\\
27 & $F_{1}^{-1}R_{1}^{-1}X_{12}F_{2}^{-1}R_{2}$\\
28 & $F_{1}^{-1}R_{1}F_{2}^{-1}X_{23}^{-1}$\\
 \hline
\end{tabular}}
\end{table*}

For the multi-qutrit system($n>3$), it is not easy to get all the
sets of MUBs, and it is even more difficult to get the explicit
decompositions of the optimal set of MUBs. Nevertheless, the results
for the two-qutrit and three-qutrit cases have provided the
experimentalists valuable references.

\section{CONCLUSION}\label{conclusion}
We have explicitly presented an optimal tomographic scheme for the
single-qutrit states, two-qutrit states and three-qutrit states
based on the MUMs. Because the MUBs based state reconstruction
process is free of information waste, the minimal number of required
conditional operations are needed. So we call our qutrits
tomographic scheme the optimal one. Here, we explicitly decompose
each measurement into several basic single- and two-qutrit
operations. Furthermore, all these basic operations have been proven
implementable \cite{ABR}. The physical complexity of a set of MUBs
also has been calculated, which is an important threshold in
experiment. We hope these decompositions can help the experimental
scientists to realize the most economical reconstruction of  quantum
states in qutrits system in Lab.

\textbf{Acknowledgments.} This work is supported by National Natural
Science Foundation of China (NSFC) under Grants No. 10704001 and No.
61073048, the Key Project of Chinese Ministry of
Education.(No.210092), the Key Program of the Education Department
of Anhui Province under Grants No. KJ2008A28ZC, No. KJ2009A048Z, No.
2010SQRL153ZD, and No. KJ2010A287, the Talent Foundation of Anhui
University, the personnel department of Anhui province, and Anhui
Key Laboratory of Information Materials and Devices (Anhui
University).


\begin{thebibliography}{99}

\bibitem{Dariano03} G. M. D'Ariano, M. G. A. Paris, and M. F. Sacchi, Advances in Imaging and Electron Physics 128,
205 (2003).
\bibitem{Dariano04} G. M. D'Ariano, M. G. A. Paris, and M. F. Sacchi, Lecture Notes in
Physics, vol. 649, Springer- Verlag, Berlin, 7 (2004).
\bibitem{GCStokes} G. G. Stokes, Trans. Cambridge Philos. Soc. 9,
399(1852).
\bibitem{JR} J. \v{R}eh\'{a}\v{c}ek \emph{et al.}, Phys. Rev. A 70, 052321 (2004).
\bibitem{Ling} A. Ling \emph{et al.}, Phys. Rev. A 74, 022309 (2006).
\bibitem{YIB1} Y. I. Bogdanov \emph{et al.}, JETP Letters, 78
(6), 352 (2003).
\bibitem{YIB} Y. I. Bogdanov \emph{et al.}, Phys. Rev. A 70, 042303 (2004).
\bibitem{PAI} P. A. Ivanov and N. V. Vitanov, Opt. Commun. 264, 368
(2006).
\bibitem{EVM} E. V. Moreva \emph{et al.}, arxiv:0811.1927v2 (2008).
\bibitem{GT} G. Taguchi \emph{et al.}, Phys. Rev. A 80, 062102 (2009).
\bibitem{AAA} A. Allevi \emph{et al.}, Phys. Rev. A 80, 022114 (2009).
\bibitem{RTT} R. T. Thew \emph{et al.}, Phys.
Rev. A 66, 012303 (2002).
\bibitem{IDI} I. D. Ivanovi\'{c}, J. Phys. A: Math. Gen, 14, 3241 (1981).
\bibitem{WKW} W. K. Wootters, B. D. Fields, Ann. Phys. (NY) 191, 363 (1989).
\bibitem{SB} S. Bandyopadhyay \emph{et al.}, Algorithmica, 34, 512
(2002).
\bibitem{Revzen} M. Revzen,  arXiv:0912.5433v1(2009).
\bibitem{JL} J. Lawrence \emph{et al.}, Phys. Rev. A 65, 032320 (2002).
\bibitem{SC} S. Chaturvedi, Phys. Rev. A 65, 044301 (2002).
\bibitem{AK} A. Klappenecker and M. R\"{o}tteler, 7th International Conference on Finite Fields and Applications,
Fq7, Lecture Notes in Computer Science 2948, 262 (2004).
\bibitem{ABK} A. B. Klimov \emph{et al.}, J. Phys. A 38, 2747 (2005).
\bibitem{JLR} J. L. Romero \emph{et al.}, Phys. Rev. A 72, 062310 (2005).
\bibitem{MP} M. Planat and H. Rosu, Eur. Phys. J. D 36, 133 (2005).
\bibitem{TD} T. Durt, J. Phys. A 38, 5267 (2005).
\bibitem{AOP} A. O. Pittenger and M. H. Rubin, J. Phys. A 38, 6005
(2005).
\bibitem{ABC} A. B. Klimov \emph{et al.}, J. Phys. A 39, 14471 (2006).
\bibitem{WW} W. K. Wootters, Found. Phys. 36, 112 (2006).
\bibitem{AJS} A. J. Scott, J. Phys. A 39, 13507 (2006).
\bibitem{ABJ} A. B. Klimov \emph{et al.}, J. Phys. A 40, 3987 (2007).
\bibitem{IB} I. Bengtsson \emph{et al.}, J. Math.
Phys. 48, 052106 (2007).
\bibitem{MAJ} M. A. Jafarizadeh \emph{et al.}, arXiv:0801.3100v1 (2008).
\bibitem{SBS} S. Brierley and S. Weigert, Phys. Rev. A 78, 042312 (2008).
\bibitem{MRK} M. R. Kibler, J. Phys. A 42, 353001 (2009).
\bibitem{PJ}  P. Jaming \emph{et al.}, arXiv:0902.0882v2 (2009)
\bibitem{SBr}  S. Brierley, S. Weigert, I. Bengtsson, arXiv:0907.4097 (2009).
\bibitem{AA} A. Ambainis, arXiv:0909.3720v1 (2009).
\bibitem{DMA}  D. M. Appleby, arXiv:0909.5233v1 (2009).
\bibitem{PRA 77} A. B. Klimov \emph{et al.}, Phys. Rev. A 77, 060303(R) (2008).
\bibitem{Bruss} D. Bruss, C. Macchiavello, Phys. Rev. Lett. 88, 127901
(2002).
\bibitem{Cerf} N. J. Cerf \emph{et al.}, Phys. Rev. Lett. 88, 127902
(2002).
\bibitem{James} D. F. V. James \emph{et al.},
Phys. Rev. A 64, 052312 (2001).
\bibitem{CM} C. M. Caves and G. J. Milburn, Opt. Commun. 179, 439
(2000).
\bibitem{ABR} A. B. Klimov \emph{et al.}, Phys. Rev. A 67, 062313 (2003).
\bibitem{MR} M. Riebe \emph{et al.}, Phys. Rev. Lett. 97,
220407 (2006).
\bibitem{012302} J. Lawrence, Phys. Rev. A 70, 012302
(2004).


\end{thebibliography}
\end{document}